# Formal Statement of the Decision-making Support Problem in the Management of Municipal Social and Economic Development

Anatoly Sidorov    Maria Shishanina
Tomsk State University of Control Systems and
Radioelectronics, 40 Lenina Prospect, Tomsk,
634050, Russian Federation
anatolii.a.sidorov@tusur.ru    mariia.a.shishanina@tusur.ru

*Abstract*—This article deals with the process of managing the social and economic development of municipal formations. It highlights characteristics and key issues that arise during management at the municipal level. In order to minimize the impact of the described issues, it is suggested to consider municipal social and economic development as a semistructured system which is modelled using a semantic network. As a result, it is concluded that a rating of indicators for assessing social and economic development needs to be created in order to determine the effectiveness and correlation with the targeted indicators.

I. INTRODUCTION

The theory and practice of the state and municipal administration are under active development, which is confirmed by inconsistencies between the legal and financial-economic situation in the territories of various levels [1], as well as by the uncertainty of the role and place of regions and municipal formations (hereinafter referred to as MF) in the current management system (including the area of social and economic development (hereinafter referred to as SED) [2]). As a result, the developed situation leads to the practical absence of territorial cooperation and ineffective usage of both budgetary and other funds [3-8]. Another difficulty is an acute differentiation between the regions of the Russian Federation (some territories show a low level of economic and financial potential, which prevents them from independent management of their development without additional assistance).

II. CHARACTERISTICS AND KEY ISSUES OF SOCIAL AND ECONOMIC DEVELOPMENT OF TERRITORIES

The SED of territories of any level is unthinkable without planning and forecasting activities which are represented by a set of strategic documents developed by bodies of authority and management in accordance with Article 11 of Federal Law of 28 June 2014, No. 172-FZ "About strategic planning in the Russian Federation".

As a result, the standing practice of strategizing has led to a number of systemic issues at the level of municipal formations:

- the prior use of administrative management methods;
- insufficient methodological support on the part of the regional authorities;
- lack of necessary competencies in the area of strategic and project management on the part of officials of different levels (especially in MF);
- strong financial dependence due to the fact that municipalities are usually deprived of additional sources of income;
- lack of an effective mechanism for interaction between local government bodies and the local community, business, and other parties concerned about the development.

Based on the understanding that the strategic planning system at the level of the MF needs to be reconsidered, which is practically impossible to implement at the municipal level, excluding the federal and regional levels, there is an objective need for increasing the level of scientific methodological relevancy of the management decisions of the bodies of authority and management made during planning, forecasting and evaluating the effectiveness of the overall social and economic development of the territory.

III. MANAGEMENT OF MUNICIPAL SOCIAL AND ECONOMIC DEVELOPMENT

The process of managing SED at the level of the MF can be shown as a functional model which represents the steps of developing basic documents (see fig. 1).

As a result, the functional model shown in figure 2 represents quite clearly the process of managing SED at the municipal level, but the issue of the content of the blocks highlighted in the model is not trivial. Consequently, it is suggested to determine the position of strategic planning documents within the organization of the system of managing SED of the MF which will be used as a link between the stages and types of strategic planning within the functional model.

Let $E = \{e_i\}$, $F = \{f_j\}$ be the sets of strategic planning procedures and strategic planning system parts, respectively. Then, the series ratio of $E$ and $F$ will allow obtaining details of the analysed sets of the strategic planning system at the municipal level. As a result of each combination of $\{e_i, f_j\}$, the corresponding strategic documents can be put together (see table 1).

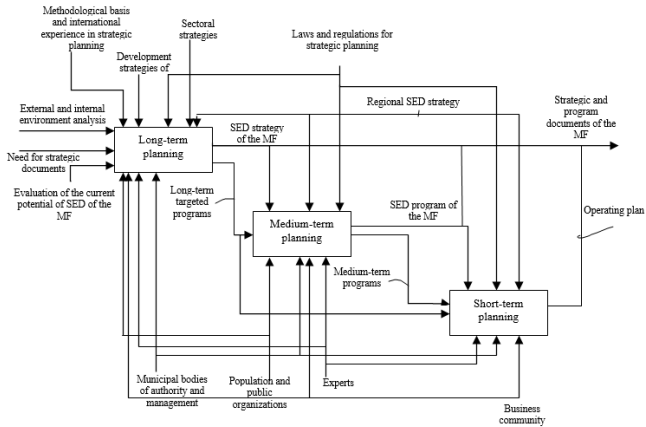

Fig. 1. Functional model of managing SED at the municipal level.

Table 1. Strategic documents of the municipal level.

| Stage of strategic planning | Type of strategic planning | | |
|---|---|---|---|
| | Long-term planning | Medium-term planning | Short-term planning |
| Target setting | - strategy for social and economic development of a municipal formation | - | - |
| Forecasting | - long-term forecast of municipal social and economic development<br>- long-term budget projection for the municipal formation | - medium-term forecast of municipal social and economic development | - |
| Planning and programming | - | - municipal programs | - plan of measures for the implementation of the municipal social and economic development strategy |

It is worth noting that the structure and content of strategic planning documents at the level of the MF are not directly regulated by the legislation. On the one hand, this aspect causes an increase in the creative initiative on the part of local authorities, while on the other — an insufficient analysis of basic documents due to the lack of necessary competencies on the part of persons making decisions in the area of SED of the MF. The latter can have a great impact if the necessary information and methodological explanations are not given at the state level since the strategic documents of municipal formations in the region will differ from each other.

## IV. STATEMENT OF THE TASK OF DECISION SUPPORT IN THE MANAGEMENT OF MUNICIPAL SOCIAL AND ECONOMIC DEVELOPMENT

The SED of the MF is a semistructured system which, according to [9], should be planned using semantic networks (see figure 2).

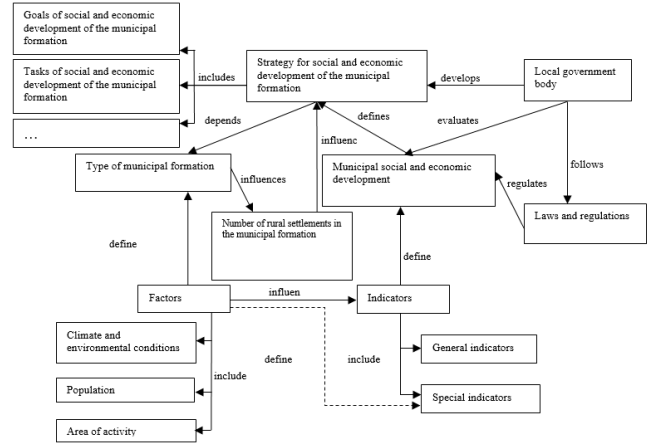

Fig. 2. Semantic network of managing SED of the MF.

Based on the general semantic network shown in Figure 3, it is worth noting that the SED strategy of the MF depends on the type of the MF, its current level of SED, as well as on the number of rural settlements. The current level of SED is meanwhile evaluated using a number of indicators that are determined by a decision-maker (hereinafter referred to as the DM). However, the list of indicators is usually transferred from the state level to the municipal one. It makes sense that the set of indicators of the SED evaluation are interrelated, which is advisable to take into account when managing SED of the MF.

Consequently, the DM should select the most significant indicators from the list and thus set the priority of the impact on their manifestation. In light of the above, the problem of evaluating the indicator rating can be represented as the following sequence: $BP = <Z, RK, \mathbf{XR} | P, \mathbf{R}>$.

Given that:

$Z = \{z_i\}$, $i = \overline{1,n}$ — a set of indicators of SED of the MF.

$RK = \{rk_x\}$, $x = \overline{1, nF}$ — a set of decision functions that determine the criticality level of the indicator depending on the evaluation of an impact on the targeted indicators.

$\mathbf{XR} = (xr_i)$, $i = \overline{1, \partial}$ — a vector of qualitative or quantitative standard (permitted) values of the indicator.

We need to find:

$P = \{o, (f_i), (k_i)\}$ — a set of indicators of the evaluation of SED of the MF, where $o$ is an estimate of the indicator relevance, $f_i$ is a vector of expert estimates of an impact of the indicator on the $i$-th targeted indicator, $k_i$ is the criticality level of the indicator with respect to the $i$-th targeted indicator.

$R_j = (r_i)$ — the rating of the indicators $z_i$ with respect to the $i$-th targeted indicator.

The expert estimates are presented according to the following rule:

$$F = \begin{cases} [-1,0), & \text{if an increase of the } i\text{-th indicator leads to a decrease of the } j\text{-th indicator;} \\ 0, & \text{if the } i\text{-th indicator has no impact on the } j\text{-th indicator (no connection);} \\ (0,1], & \text{if an increase of the } i\text{-th indicator leads to an increase of the } j\text{-th indicator.} \end{cases}$$

## V. Conclusions

The result is that determining the rating of indicators will allow the DM to find the most crucial ones that affect the targeted indicators (with respect to SED of the MF, the quality of life will be the targeted indicator, as it is described as a first-priority indicator by regulatory documents). It is logical to assume that the implementation of certain measures of SED of the MF will affect the indicators, including the targeted ones. Thus, the bodies of authority and management can determine measures that allow achieving the stated strategic goals and objectives for the development of a particular territory. At the same time, the set of indicators should be considered dynamically by taking into account their correlation.

## References

[1] Khayrullov D S and Eremeev L M 2012 Stability issues of social and economic development of the region *Bulletin of the Kazan State Agrarian University* **1** 73-6
[2] Nikitenko S M and Goosen E V 2017 Socio-economic development of territories based on the principles of public-private partnership in the sphere of comprehensive mineral exploration *IOP Conference Series: Earth and Environmental Science* **84(1)** 012013
[3] Raymbaev C K, Kulueva C, Giyazov A, Bezrukov B A and Bezrukova T L 2017 Concept of Innovational Development of Entrepreneurial Potential of Small Enterprises *Contributions to Economics* **9783319454610** 143-50
[4] Shishanina M A 2017 Decision support system for socio-economic development of rural settlements: general provisions *Proc. Int. Conf. TUSUR Scientific Session* part 8 (Tomsk, Russia: V-Spektr) pp 210-2
[5] Vetrov G Yu 2009 *Management of municipal economic development* (Moscow, Russia: Foundation of The Institute of Urban Economics)
[6] Oreshnikov V V and A G Ataeva 2017 Methodological aspects of assessing the level of social and economic development of municipalities from the position of their territorial transformation *Bulletin of the Belgorod University of Cooperation, Economics and Law* **6(67)** 130-46
[7] Sidorov A A and Silich M P 2008 Methodological approaches to assessing social and economic development of municipalities *Tomsk Polytechnic University Newsletter* **6** 38-44
[8] Maksimov V I 2005 Structure-objective analysis of socio-economic situations development *Management Problems* **3** 30-8
[9] Sidorov A A and Shishanina M A 2017 Semantic network as a tool for determining the process of management of municipal social and economic development *Electronic means and control systems* **1-2** 188-92